\documentstyle[12pt]{article}
\input amssym.def
\def\openoneh{\leavevmode\hbox{\small1\kern-3.8pt\normalsize1}_{\cal
H}}
\newcommand{\Trh}[1]{{\mathrm{Tr}}_{{\cal{H} }}\left( #1 \right)}
\newcommand{\Tr}[1]{{\mathrm{Tr}}_{{\cal{H} }}\left( #1 \right)}
\newcommand{\Avg}[2]{\left< #2 \right>_{#1}}
\newcommand{\T}{{\mathrm{T}}}
\newcommand{\Space}{\;\;\;\;\;}
\newcommand{\dfid}{D_{\mathrm{fid}}}

\input psfig

\begin{document}
\begin{titlepage}
\title{Security against eavesdropping in quantum cryptography}
\author{ Norbert L\"utkenhaus and Stephen M.~Barnett\\University of
Strathclyde, Glasgow G4 0NG, Scotland\\ EMail: \{norbert,
steve\}@phys.strath.ac.uk}
\date{25.~September 1996}
\end{titlepage}
\maketitle

\setcounter{page}{1}
\section{Introduction to quantum cryptography}
Quantum cryptography is a method for providing two parties who want to
communicate securely with a secret key to be used in established
protocols of classical cryptography. For more reviews of this topic
see \cite{hughes95a,phoenix93a,phoenix95a}. Bennett and Brassard
showed that it is possible, at least ideally, to create a secret key,
shared by sender and receiver, without both parties sharing any secret
beforehand. We refer to this protocol as the {\em BB84}
protocol. \cite{bennett84a} To achieve this goal, sender and receiver
are linked by two channels. The first channel is a public channel. The
information distributed on it is available to both parties {\em and}
to a potential eavesdropper. To demonstrate the principle of quantum
cryptography we assume that the signals on this channel can not be
changed by third parties. The second channel is a channel with strong
quantum features. An eavesdropper can interact with the signal in an
effort to extract information about the signals. The signal states are
chosen in such a way that there is always, on average, a back reaction
onto the signal states. We assume the quantum channel to be noiseless
and perfect so that the back reaction of the eavesdropper's activity
manifests itself as an induced error rate in the signal transmission.

The BB84 protocol uses the polarisation states of single photons as
signal states. The signal states are, for example, linear vertical or
horizontal polarised photons or right or left circular polarised
photons. The sender sends a sequence of single photons with a
polarisation chosen randomly from the four given ones. The receiver
uses randomly one out of two given polarisation analysers for each
signal photon. One of the analysers distinguishes between the two
linear polarisations, the other between the circular
polarisations. Therefore the sequence of signals contains two types of
transmissions.  In the first type the photon is prepared in a
polarisation state which the polarisation analyser, chosen by the
receiver, is able to distinguish unambiguously. An example is that a
horizontal polarised photon is sent and the receiver chooses to use
the linear polarisation analyser.  Signals of this type will be
refered to as deterministic signals since, the outcome of the
polarisation measurement is fully determined by the state of the
signal photon. The remaining signals are non-deterministic signals. An
example for this is a horizontal linear polarised photon which
triggers with equal probability the outcome ``right circular'' and
``left circular'' in the polarisation analyser distinguishing in the
circular polarisation basis. Sender and receiver can distinguish
between deterministic and non-deterministic signals using the public
channel without giving away any information about the specific
signal. They just compare the polarisation basis of the signal and the
measuring polarisation analyser, both of which can be ``linear'' or
``circular''. The signal sequence of the deterministic bits can then
be transformed into a binary key by assigning ``0'' for linear
horizontal or right circular polarised photons and ``1'' for the
remaining linear vertical or left circular polarised photons.

In this idealisation the security of quantum cryptography is given by
the fact that the four polarisation states are not four orthogonal
states and so there exists no quantum non-demolition measurement which
could distinguish between them.  Even more importantly, each attempt
to distinguish between any of the states will change, in average, the
states of the signals. This state-change destroys the perfect
correlation between the signal and the measurement outcomes for the
deterministic signals. A test checking this correlation will reveal
any attempt at eavesdropping. If the test shows that the correlation
is still perfect then sender and receiver can be sure that there was
no eavesdropping attack and their shared binary key is perfectly
secret.

In the practical realisation of quantum cryptography we face two
specific problems. The public channel needs to be implemented in such
a way that sender and receiver can ensure that the messages being
received are really coming from each other. This is the problem of
{\em authentication} for which various techniques are in use. In
general, however, there will be the need for the two parties to share
a limited amount of secret knowledge, for example in form of a secret
key, before the authentication can take place. Quantum cryptography
then generates a large secret key from a small secret key.

The work presented here deals with the second problem arising from the
fact that all realistic quantum channels are noisy. Therefore the
correlation between the signals and the measurement outcomes for the
deterministic signals will not be perfect. Noise has the same effect
on the signals as the activity of an eavesdropper. It is therefore
necessary to think of all state change of the signals to be due to
eavesdropping activity.  It is intuitively clear that an eavesdropper
can only have gained a small amount of information on the key if the
correlation tested by sender and receiver are still strong, that is,
if there are only a few transmission errors for the deterministic
signals. One can hope to give a bound on the eavesdropper's Shannon
information as a function of the error rate in the deterministic
signals. Such bounds have been obtained assuming that the eavesdropper
is restricted to von Neumann measurements only \cite{huttner94a} or to
a restricted class of more general measurements \cite{ekert94a}. Here
we present a sharp bound \cite{nl96a,nl96b} on the Shannon information
of an eavesdropper which is valid for all eavesdropping attacks which
access each signal photon independently of each other. It therefore
does not include coherent attacks in which the product state of all
signal photons is attacked. The sharp bound does not take into account
that an eavesdropper can make use of the later acquired knowledge
about the polarisation basis of the signal photons to {\em change} the
measurement of the signal. However, we are able to give a rough upper
bound for this situation. The reason that this is possible is that the
eavesdropper has to decide how to interact with the signal states
before he acquires the additional knowledge about the polarisation
basis.

\section{Generalised measurements}
The key input to derive the bounds given in this paper is the most
general description of a measurement on a given system. Any
measurement can be described by a set of operators $A_l$ defined on
the Hilbert space $\cal H$ of the system the measurement is performed
on. The only restriction on the operators is that
\begin{equation}
\label{sumstuff}
\sum_{l \in K} A^\dagger_l A_l = \openoneh \; .
\end{equation}
where $K$ is some finite or countable infinite index set. The link
between these operators and a measurement is given by the following
formulas which describe the probabilities that a particular outcome of
a measurement is triggered, and which give the final state of the
measured system after that outcome was registered. For the sake of
simplicity we assume the set of outcomes to be discrete. The
probability that the outcome $k$ is triggered by an input state with
density matrix $\rho$ is given by
\begin{equation}
p_k = \Trh{\rho \sum_{l \in K_k} A^\dagger_l A_l} \;
\end{equation}
 where the $K_k$ are disjunct subsets of $K$ with $K = \bigcup_{k}
 K_k$. The final density matrix $\tilde{\rho}^{(k)}$ of the selected
 states belonging to this outcome is given by
\begin{equation}
\label{discretselective}
\tilde{\rho}^{(k)} = \frac{ \sum_{l \in K_k} A_l \rho A^\dagger_l}
{\Trh{\rho \sum_{l \in K_k} A^\dagger_l A_l}} \; .
\end{equation}
The density matrix of the final state, which does not select any
states, but describes the whole ensemble for all outcomes is given by
\begin{equation}
\label{discretnonselective}
\tilde{\rho} = \sum_k p_k \tilde{\rho}^{(k)} = \sum_{l \in K} A_l \rho
A^\dagger_l \; .
\end{equation}
It is important to choose the correct Hilbert space $\cal H $ to
describe the measurement. To describe a spin measurement of an
electron and the back reaction onto that spin we will choose $\cal H$
to be the Hilbert space of the spin of electron. If we are interested
in the position or momentum of the electron as well we have to add the
Hilbert space of spatial modes.  It is a bit less obvious in quantum
optics. If $\cal H$ is the one-photon polarisation Hilbert space then
we implicitly assume that precisely one photon remains after the
measurement. An eavesdropper may absorb the photon and so the final
state is the vacuum.  Therefore the adequate Hilbert space is the full
Hilbert space of a light mode plus the polarisation degree of
freedom. It turns out that we can derive the bounds presented here by
restricting the Hilbert space to that of a single photon and later
generalise it to the full Hilbert space.

\section{Estimate of the Shannon information}
The relevant expression for the Shannon information per signal is
given with the help of the function $h(x) = -x \log x$, where $\log$
refers to basis 2, as
\begin{equation}
I = \sum_{\Psi = 0,1} h\left[ p(\Psi) \right] + \sum_{ \alpha = \circ,
+ \atop k} h\left[ p(k_\alpha) \right] - \sum_{\Psi = 0,1 \atop k,
\alpha} h\left[ p(\Psi, k_\alpha) \right] \; .
\end{equation}
An eavesdropper gains this Shannon information on the binary key whose
signals are given by $\Psi = 0,1$ when he learns, from communication
on the public channel, the polarisation basis $\alpha =\circ, +$
(linear or circular) used for each signal, and registers the outcome
$k$ on the measurement apparatus triggered by each signal. The
probabilities that a ``0'' or a ``1'' is sent are denoted by
$p(\Psi)$, the probability that outcome $k$ is triggered by a photon
prepared as basis state of the linear or circular polarisation basis
is written as $p(k_\alpha)$, and the joint probability distribution
for both events is $p(\Psi, k_\alpha)$.

We would like to give an upper bound on this Shannon information as a
function of the {\em measured} disturbance of the quantum channel
given as the error rate in the deterministic signals. This error rate
is basically the fidelity measure $\dfid$ of the channel, given by
\begin{equation}
\dfid = 1- \frac{1}{4} \sum_{j=1}^4 \Tr{ \rho_i \tilde{\rho}_i}.
\end{equation}
The definition of $\dfid$ assumes here the use of the one photon
polarisation Hilbert space.  The expression $\Tr{ \rho_i
\tilde{\rho}_i}$ is the overlap between the input state $\rho_i$,
which is one of the four signal states, and the final state
$\tilde{\rho}_i$ of the eavesdropper's measurement performed on this
input state $\rho_i$. This can be interpreted as the probability that
the final state is still recognised as the initial state in apparatus
of the receiver. Then $\dfid$ is the error rate averaged over the four
signal states.

It can be shown \cite{nl96b} that the optimal eavesdropping strategy
can be given by operators $A_k$ which can be described by real
matrices in a representation for which the signal states are real
density matrices.  Each operator $A_k$, which can always be expressed
in the form
\begin{equation}
\label{symmetricorth}
A_k = \sqrt{a_k} O_k + \left( \sqrt{b_k} - \sqrt{a_k} \right) O_k P_k
\end{equation}
with real, positive numbers $a_k$ and $b_k$, ($b_k \geq a_k$),
projection operators $P_k$ and orthogonal operators $O_k$. The optimal
strategy satisfies the symmetry that for each such operator the set of
operators $A_k$ contains as well the operator
\begin{equation}
\label{symmetrictilde}
\tilde{A}_k = \sqrt{a_k} O_k + \left( \sqrt{b}_k - \sqrt{a_k} \right)
O_k \overline{P} \;
\end{equation}
which employs the orthogonal complement $\overline{P} = \openoneh
-P_k$.  The optimal eavesdropping strategy associates a measurement
outcome with each of the operators $A_k$ separately so that we do not
need to employ any partitions $K^{(k)}$. For such an eavesdropping
strategy the measured disturbance $\dfid$ is given by
\begin{eqnarray}
\dfid & = & \sum_i \Bigg[ \frac{1}{4} \Trh{ \rho_i E_i} -\sum_{k \in
 K}\Bigg( \frac{1}{4} \sqrt{a_k b_k} \Tr{ O_k \rho_i O_k^\T E_i}
 \nonumber \\ & & + \frac{1}{4} \frac{\left( \sqrt{b_k} - \sqrt{a_k}
 \right)^2}{2} \left(\Tr{O_k P_k \rho_i P_k O_k^\T E_i} + \Tr{ O_k
 \overline{P}_k \rho_i \overline{P}_k O_k^\T E_i}\right)
 \Bigg)\Bigg]\; .  \nonumber
\end{eqnarray}
Here $E_i$ is the projection operator describing the effect of the
polarisation analyser. For the one photon space we have $E_i =
\rho_i$.  The Shannon information is given by the expression
\begin{equation}
\label{Icd}
\begin{array}{rlll}
 I =&\multicolumn{3}{l}{ \sum_k \frac{a_k + b_k}{2} \bigg[1 -
 \log(1+\eta_k^2) +} \nonumber \\ &
 &\multicolumn{2}{l}{\frac{1}{2(1+\eta_k^2)} \bigg\{ (\eta_k^2 + c_k -
 \eta_k^2 c_k) \log (\eta_k^2 + c_k - \eta_k^2 c_k)} \nonumber \\ &
 &\hspace{0.5cm} & + (1-c_k + \eta_k^2 c_k) \log(1-c_k + \eta_k^2
 c_k)\nonumber \\ & & &+ (\eta_k^2 + d_k - \eta_k^2 d_k) \log(\eta_k^2
 + d_k - \eta_k^2 d_k) \nonumber\\ & & &+ (1-d_k + \eta_k^2 d_k)
 \log(1-d_k + \eta_k^2 d_k) \bigg\} \bigg] \; .\nonumber
\end{array}
\end{equation}
In this expression we used the definitions of the overlaps $c_k =
\Trh{\rho_1 P_k}$ and $d_k = \Trh{\rho_3 P_k}$ and of the
characteristic parameters $\eta_k =\sqrt{\frac{a_k}{b_k}}$ with $\eta
\in [0,1]$.  For $\eta_k=1$ an operator $A_k$ takes the
characteristics of the identity operator, which corresponds to
non-interference of the eavesdropper, and for $\eta_k = 0$ the
eavesdropping strategy tends to a von Neumann projection
measurement. The overlaps are restricted by the inequality
\begin{equation}
\label{circlestuff}
(d_k-\frac{1}{2})^2 + (c_k - \frac{1}{2})^2 \leq \frac{1}{4} \; .
\end{equation}

We can find the optimal choice of orthogonal operators $O_k$ and
$P_k$. The optimal choices are given in a later section. As a result
we find now for the disturbance the inequality
\begin{equation}
 \label{destimateab} \dfid \geq \sum_{k \in K} \frac{a_k + b_k}{2}
\frac{1}{4} \frac{\left( 1 - \eta_k \right)^2}{1+ \eta_k^2} \; .
\end{equation}
Note that the condition (\ref{sumstuff}) implies that
\begin{equation}
\sum_k \frac{a_k + b_k}{2} = 1
\end{equation}
so that the expressions $\frac{a_k + b_k}{2}$ have the property of a
probability.  The Shannon information can be estimated by
\begin{equation}
\label{shannon_1}
I \leq \sum_k \frac{a_k + b_k}{2} \frac{1}{2} \left( 1 -
\log(1+\eta_k^2) + \frac{\eta_k^2}{1 + \eta_k^2} \log \eta_k^2 \right)
\; .
\end{equation}

\begin{figure}[htb]
\centerline{\psfig{figure=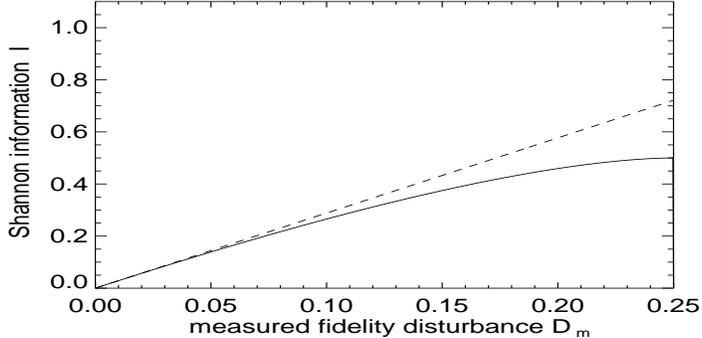,width=4in,height=2in}}
\parbox{14cm}{\caption{ \em The bound on the Shannon information in
the sharp (continuous line) and the linear bound (dashed line) as a
function of the measured disturbance $D_m$.}}
\label{shannon_d}
\end{figure}

It can be shown that the optimal choice of the characteristic
parameters $\eta_k$ is for them to take the same value
$\tilde{\eta}$. The proof uses variation methods. Then we find the
inequalities
\begin{eqnarray}
 D_{fid} & \geq & \frac{1}{4} \frac{\left( 1 - \tilde{\eta}
\right)^2}{1 + \tilde{\eta}^2} \\ I & \leq &\frac{1}{2}\left( 1 -
\log(1+\tilde{\eta}^2) + \frac{\tilde{\eta}^2}{1 + \tilde{\eta}^2}
\log \tilde{\eta}^2 \right) \; .
\end{eqnarray}
If we actually measure the average error rate $\dfid$ and find the
value $D_m$ we can bound the value of $\tilde{\eta}$ by
\begin{equation}
\label{measuredeps}
\tilde{\eta} \geq \overline{\eta} := \left\{
\begin{array}{ll}
\frac{1- 2 \sqrt{2} \sqrt{\left( 1 - 2 D_m \right) D_m}}{1 - 4 D_m} &
 \Space D_m \leq \frac{1}{4} \\ 0 & \Space D_m \geq \frac{1}{4}
\end{array}
\right.  \;
\end{equation}
 which leads to the bound of the eavesdropper's Shannon information
 \cite{nl96a,nl96b} as
\begin{equation}
\label{shannonsharp}
I_S \leq \frac{1}{2} \left(1 - \log(1+\overline{\eta}^2) +
\frac{\overline{\eta}^2}{1 + \overline{\eta}^2} \log \overline{\eta}^2
\right) \; .
\end{equation}
  It can be shown that this bound can be further estimated by the
  linear bound
\begin{equation}
I \leq \frac{2}{\ln 2} D_m
\end{equation}
where $\ln 2$ is the natural logarithm of $2$. For small $D_m$ this
bound is nearly as good as the bound (\ref{shannonsharp}) which will
later be shown to be sharp.  The sharp bound and the linear
approximation are plotted in figure \ref{shannon_d} as a function of
the measured fidelity disturbance.
Typical values for experimental realisations using the BB84 protocol
achieve an error rate of 4 \% for 30 km or 1.5 \% for 10 km distance
between sender and receiver.

\section{Privacy amplification}
For the purpose of secret communication the amount of Shannon
information possibly leaked to an eavesdropper according to the
previous estimate is far to high. However, making use of the technique
of {\em privacy amplification} \cite{bennett95a} it is possible to
reduce Eve's total amount of Shannon information on the remaining
key. For that the key has to be shortened using {\em hash
functions}. The characteristic quantity for the fraction by which the
key has to be shortened is the parameter $\tau_1$. If we shorten the
key by the fraction $\tau_1$ then the eavesdropper is left at most
with a Shannon information of 1 bit on the {\em whole} key. Each bit
by which the key is is shortened additionally decreases this remaining
Shannon information exponentially. The parameter $\tau_1$ can be
expressed with the help of the {\em collision probability}
$\Avg{y}{p_c(y)}$ as
\begin{equation}
\label{taudef}
\tau_1 = 1 + \log \Avg{y}{p_c(y)}^\frac{1}{N} \;.
\end{equation}
The collision probability $\Avg{y}{p_c(y)} = \sum_x p(x|y)^2 $ refers
to the probability distribution $p(x|y)$ over all possible signal
string $x$, conditioned on the event that the eavesdropper measured a
particular string of measurement results $y$.

Before they can apply the technique of privacy amplification, the
sender and the receiver have to perform some type of error correction
on their shared key. We assume that this process can be done without
the eavesdropper gaining any additional knowledge and without the
creation of any correlations between the signals.  A possible
realisation would be to use block parity comparison where the compared
parity bit is encoded using some short shared secret key from the same
source which gives the key used in the authentication of the public
channel. In this case the collision probability is given by
\begin{equation}
\left< p_c(y) \right>^{\frac{1}{n}} = \sum_{k, \alpha, \psi} \frac{
p^{(c)}\left( \psi, k_\alpha \right) ^2 }{p^{(c)}\left( k_\alpha
\right)} \; ,
\end{equation}
where the probabilities $p^{(c)}\left( \psi, k_\alpha \right)$ and
$p^{(c)}\left( k_\alpha \right)$ now refer to the corrected key, and
it takes only those signal transmissions which were correctly received
into account. The joint probability distribution $p^{(c)}\left( \psi,
k_\alpha \right)$ is given, with a normalisation constant $C$, by
\begin{equation}
\label{pcorrect}
p^{(c)}\left( \psi, k_\alpha \right) =\frac{1}{C} \Trh{A_k
\rho_{\psi_\alpha} A_k^\dagger \rho_{\psi_\alpha}} \; .
\end{equation}

It is again possible to show general properties of the  $A_k$
which lead to an optimal information gain by the eavesdropper, along
with minimal disturbance of the signal transmission. The optimal
 $A_k$ can be shown to be real (in the real representation of
the signal states) and to consist of symmetric or anti-symmetric
matrices. The symmetric matrices can have eigenvalues of different or
of the same sign so that they can be written as
\begin{equation}
\label{symmetric}
A^{(\pm)}_k= \sqrt{a_k} \openoneh - \left( \sqrt{a_k} \pm \sqrt{b_k}
\right) P_k
\end{equation}
with the $a_k$ and $b_k$ satisfying, as before, $b_k \geq a_k \geq 0$
and $P_k$ is a projection operator. To each such operator the set of
operators $A_k$ contains an operator
\begin{equation}
\label{symmetrictilde2}
\tilde{A}^{(\pm)}_k= \sqrt{a_k} \openoneh - \left( \sqrt{a_k} \pm
\sqrt{b_k} \right) \overline{P_k}
\end{equation}
using the orthonormal complement $\overline{P_k} = \openoneh - P_k$. I
refrain from giving to give the expressions for the collision
probability and the disturbance $\dfid$ in the general form and give
instead the forms optimised with respect to the choice of projection
operators $P_k$. At this stage they are given with the help of the
characteristic parameter $\eta_k$ which satisfies $\eta_k^2 =
\frac{a_k}{b_k}$ and takes values in the range $\eta_k \in [-1,1]$
which is in contrast to the calculations leading to the bound on the
Shannon information.  The disturbance satisfies the inequality
\begin{equation}
\label{dresultcoll}
 \dfid = \sum_{k \in K^{(\pm)}} \frac{a_k + b_k}{2} \frac{1}{4}
 \frac{\left( \eta_k - 1 \right)^2}{\eta_k^2 + 1} + \sum_{k \in
 K^{(a)}} \frac{a_k + b_k}{2} \frac{1}{4} \; . \nonumber
\end{equation}
where $K^{(\pm)}$ is the index set of the symmetric operators and
$K^{(a)}$ the index set of the anti-symmetric operators. The collision
probability is bound by
\begin{equation}
\left< p_c^{(c)}(y) \right>^{\frac{1}{n}} \leq \frac{\displaystyle
\sum_{k} \frac{a_k + b_k}{2}\frac{1}{2} \frac{1}{ 1 + \eta_k^2}
\frac{17 + 12 \eta_k + 6 \eta_k^2 + 12 \eta_k^3 + 17 \eta_k^4}{3 + 2
\eta_k + 3 \eta_k^2}}{\displaystyle \sum_{k}\frac{a_k + b_k}{2}\frac{3
+ 2 \eta_k + 3 \eta_k^2}{ 1 + \eta_k^2}} \; .
\end{equation}

\begin{figure}[htb]
\centerline{\psfig{figure=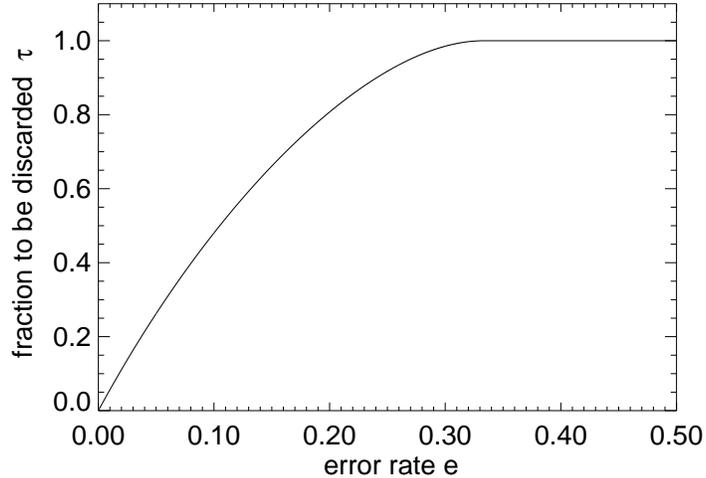,width=4in}}
\parbox{14cm}{\caption{ \em The parameter $\tau_1$ as a function of
the error rate. The error rate $e$ is equal to the disturbance measure
$\dfid$.}}
\label{tau1_c}
\end{figure}

We use again a variation method to show that the optimal eavesdropping
strategy employs characteristic parameters $\eta_k$ with the
same value $\tilde{\eta}$. Also it is  clear that it
is of disadvantage to the eavesdropper to use anti-symmetric operators
$A_k$. This leads to the estimates
\begin{equation}
\dfid \geq \frac{1}{4} \sum_{k \in K} \frac{\left( \tilde{\eta} - 1
 \right)^2}{\tilde{\eta}^2 + 1} \nonumber
\end{equation}
and
\begin{equation}
\left< p_c^{(c)}(y) \right>^{\frac{1}{N}} \leq \frac{1}{2} \frac{17 +
12 \tilde{\eta} + 6 \tilde{\eta}^2 + 12 \tilde{\eta}^3 + 17
\tilde{\eta}^4}{\left(3 + 2 \tilde{\eta} + 3 \tilde{\eta}^2\right)^2}
\; .
\end{equation}
The measured disturbance $D_m$ leads to a bound on $\tilde{\eta}$
given by
\begin{equation}
\label{etaoverlined}
\tilde{\eta} \geq \overline{\eta} := \left\{
\begin{array}{ll}
\frac{1- 2 \sqrt{2} \sqrt{\left( 1 - 2 D_m \right) D_m}}{1 - 4 D_m} &
 \Space D_m \leq \frac{1}{2} \\ -1 & \Space D_m \geq \frac{1}{2}
\end{array}
\right.  \; .
\end{equation}
This finally allows us to bound the parameter $\tau_1$ (\ref{taudef})
by the inequality \cite{nl96b}
\begin{equation}
\label{taufinal}
\tau_1 \leq \left\{
\begin{array}{ll}
\log \left( \frac{17 + 12 \overline{\eta} + 6 \overline{\eta}^2 + 12
\overline{\eta}^3 + 17 \overline{\eta}^4}{\left(3 + 2 \overline{\eta}
+ 3 \overline{\eta}^2\right)^2}\right) & \Space D_m \leq \frac{1}{3}
\\ 1 & \Space\frac{1}{3} \leq D_m \leq 1
\end{array}
\right.
\end{equation}
This bound is shown in figure \ref{tau1_c}.
Typical error rates in the BT experiment are $e \in [0.01,0.05]$ which
corresponds to $\tau_1 \in [0.05, 0.26]$.

\section{Validity of the bounds}
The derivation of the bounds presented above assumes that the
eavesdropper interacts with the signal photons but does not absorb
them. In the experimental realisation, however, an absorption of about
90 \% is observed. The validity of the bounds can be extended to
accommodate the possibility of absorption by re-defining the average
error rate as refering only to those signals where the polarisation
analyser successfully measured a signal. It can be shown that the
eavesdropper cannot increase the trade-off between information and
induced error rate by forwarding signal states to the receiver which
contain more than one photon. The basic tool for this extension of the
validity of the bounds is that one can show that each eavesdropper
strategy is equivalent to an eavesdropper strategy which results in
final states which are Fock states of fixed photon number.

\section{Delayed measurements}
\begin{figure}[htb]
\centerline{\psfig{figure=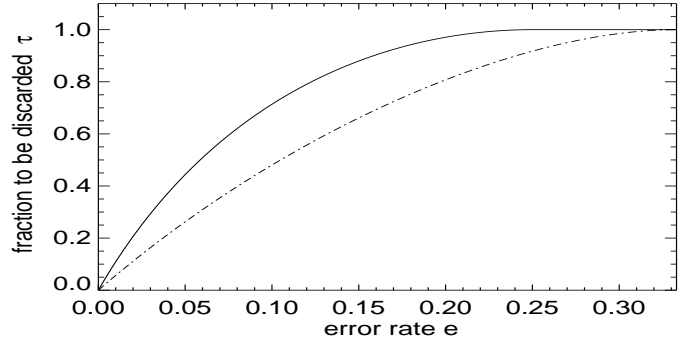,width=4in,height=2in}}
\parbox{14cm}{\caption{ \em The crude bound of the fraction $\tau_1$
of bits to be discarded during privacy amplification allowing for
delayed measurements (solid line). This is compared to the  sharp bound
 for non-delayed measurements (dashed line). }}
\label{delayed_c}
\end{figure}

The description of a delayed measurement needed here is that the
eavesdropper has effectively two eavesdropping strategies at hand: one
for each signal set of linear or circular polarisation. They are given
by the A-operators $\left\{ A_k \right\}_{k \in K}$ and $\left\{B_l
\right\}_{l \in L}$ with two index sets $K$ and $L$ which are not
necessarily of the same size. The strategies cannot be chosen
independently of each other since they must be alternative
descriptions of the quantum channel viewed as a non-selective
measurement. This means that the equality
\begin{equation}
\label{delayedrestriction}
\sum_{k \in K} A_k \rho A_k^\dagger = \sum_{l \in L} B_l \rho
B_l^\dagger
\end{equation}
must hold for all density matrices $\rho$. An example of relations
between the sets $\left\{ A_k \right\}_{k \in K}$ and $\left\{ B_l
\right\}_{l \in L}$ satisfying this equality is the choice
$
B_l = \sum_k c_{lk}A_k
$
with $\sum_k c_{lk} \overline{c}_{kn} = \delta_{ln}$. One can give a
crude estimate of the Shannon information and of the collision
probability because the disturbance is independent of the overlaps
$c_k$ and $d_k$. To give the bounds let the eavesdropper 
choose the projection operators $P_k$ of the set of operators $A_k$
and $B_k$ independently. Quantum mechanics will put some
restrictions on that relation so that the resulting bounds are no
longer sharp. Thus the eavesdropper's Shannon information may
increase by a factor  2. The collision probability  is  bounded by
\begin{equation}
\left< p_c^{(c)}(y) \right>^{\frac{1}{n}} \leq \frac{1 +
\tilde{\eta}^4}{(1 + \tilde{\eta}^2)^2}
\end{equation}
where $\tilde{\eta}$ is bounded by the measured disturbance $\dfid$ as
given in (\ref{etaoverlined}).  The resulting bound for the fraction
$\tau_1$ of bits to be discarded during privacy amplification is
plotted in figure \ref{delayed_c}.
In this estimate we can prove security against eavesdropping as long
as the error rate is less than $25$ \%.  This result is likely to  remain valid if we allow $n$-photon operations as
 in the previous chapter.  Formal proof, however, should
be postponed until a sharp bound for delayed choice eavesdropping
strategies can be given.  Clearer understanding of the restrictions
imposed by (\ref{delayedrestriction}) is essential for the derivation
of the sharp bound.

\bibliography{norbert}
\bibliographystyle{osa}

\end{document}